\documentclass[aps+-,prl,twocolumn,amsfonts,amssymb,amsmath,nofootinbib,superscriptaddress]{revtex4-2}

\usepackage{graphicx} 
\usepackage{nicematrix} 
\usepackage{svg} 

\usepackage[T1]{fontenc} 
\usepackage[english]{babel}

\usepackage{hyperref} 
\usepackage[capitalize]{cleveref} 

\newcommand{\up}[1]{\left|\uparrow,#1\right\rangle}
\newcommand{\down}[1]{\left|\downarrow,#1\right\rangle}
\newcommand{\ket}[1]{\left|#1\right\rangle}

\begin{document}
\title{Coherent dynamics of a nuclear-spin-isomer superposition}
\author{Tamar Levin}
\email{tamar.levin@weizmann.ac.il}
\author{Ziv Meir}
\email{ziv.meir@weizmann.ac.il}
\affiliation{Department of Physics of Complex Systems, Weizmann Institute of Science, Rehovot 761001, Israel}

\begin{abstract}
Preserving quantum coherence with the increase of a system's size and complexity is a major challenge.
Molecules, with their diverse sizes and complexities and many degrees of freedom, are an excellent platform for studying the transition from quantum to classical behavior. 
While most quantum-control studies of molecules focus on vibrations and rotations, we focus here on creating a quantum superposition between two nuclear-spin isomers of the same molecule.
We present a scheme that exploits an avoided crossing in the spectrum to create strong coupling between two uncoupled nuclear-spin-isomer states, hence creating an isomeric qubit.
We model our scheme using a four-level Hamiltonian and explore the coherent dynamics in the different regimes and parameters of our system.
Our four-level model and approach can be applied to other systems with a similar energy-level structure.

\end{abstract}
\maketitle
\section{Introduction}
Quantum mechanics is the underlying theory of matter, light, and their interaction. While it is essential for describing the microscopic world, a much simpler classical description is sufficient on a macroscopic scale due to environmental-induced decoherence~\cite{zurek1991decoherence}.

There has been much scientific effort to retain the quantum behavior of macroscopic systems in a well-controlled lab environment. These studies try to push further away the transition between quantum and classical behavior. A few examples are matter-wave interferometry of large-scale molecules and clusters~\cite{hornberger2012colloquium} and the entanglement of macroscopic mechanical objects~\cite{kotler2021direct}.

A recent breakthrough that garnered attention in the physics community was the laser excitation and resonance-fluorescence signal of the $^{229m}\text{Th}$ nucleus~\cite{Tiedau2024,zhang2024dawn}. This result sparked the imagination for creating a coherent quantum superposition of the $\text{Th}$ atom and its nucleus isomer with applications in dark-matter searches and precision metrology~\cite{peik2021nuclear}. 

From the many newly developed quantum hardware, molecules stand out due to their additional degrees of freedom, such as vibrations, rotations, and different isomers. Reaching the quantum regime in molecules can be attained by assembling molecules from ultracold atoms~\cite{Moses2016NewMolecules}, laser cooling~\cite{fitch2021laser}, deceleration~\cite{Segev2019CollisionsTrap}, and sympathetic cooling in ion traps~\cite{molhave2000formation,schwegler2023trapping}. Together with quantum-control techniques such as quantum logic~\cite{Schmidt2005}, the latter approach allowed for quantum-non-demolition state detection~\cite{Wolf2016, Sinhal2020Quantum-nondemolitionMolecules, Najafian2020IdentificationForcesb}, coherent manipulation~\cite{chou2017}, precision spectroscopy~\cite{Chou2020Frequency-combIonc}, and ion-molecule entanglement~\cite{Lin2020QuantumMolecule} of rotational and vibrational states in molecular ions. 

While the community is focused on molecules' rotational and vibrational degrees of freedom, the coherence between isomeric degrees of freedom remains unexplored. Here, we propose to attain quantum coherence between different isomeric degrees of freedom in molecules. Specifically, we devise a scheme to create a quantum superposition of two distinct nuclear-spin isomers of the nitrogen molecular ion. 

\section{Nuclear-spin-isomer mixing} 
The exchange symmetry profoundly affects the rotational spectrum of molecules with identical nuclei, such as homonuclear molecules. Since the total molecular wavefunction must be symmetric (antisymmetric) under the exchange of identical bosonic (fermionic) nuclei, only odd or even rotational states are allowed in the spectrum~\cite{tino2022identical}. For example, the $^{14}\textrm{N}_2^{+}$ molecular ion comprises two identical bosonic nuclei, each with nuclear spin $I_1=1$. The molecule consists of three (total) nuclear-spin isomers (NSIs), $I=0,1,2$. In the electronic ground state, the two ortho-isomers ($I=0,2$) are associated with the even rotational states, while the para-isomer ($I=1$) is associated with the odd rotational states. 

The parity selection rule ($\Delta N=2$) imposed by the exchange symmetry in homonuclear molecules forbids the mixing of adjacent rotational states by electric-field dipole coupling, protecting them from blackbody radiation. Therefore, the entire rovibrational manifold in the electronic ground state of homonuclear molecules is long-lived and appealing for applications in quantum information as qubits~\cite{Sinhal2020Quantum-nondemolitionMolecules, Najafian2020IdentificationForcesb} and in metrology as molecular clocks~\cite{Schiller2014,leung2023terahertz, madge2024prospects}.    

\begin{figure}
    \centering
    \includegraphics[width=\linewidth]{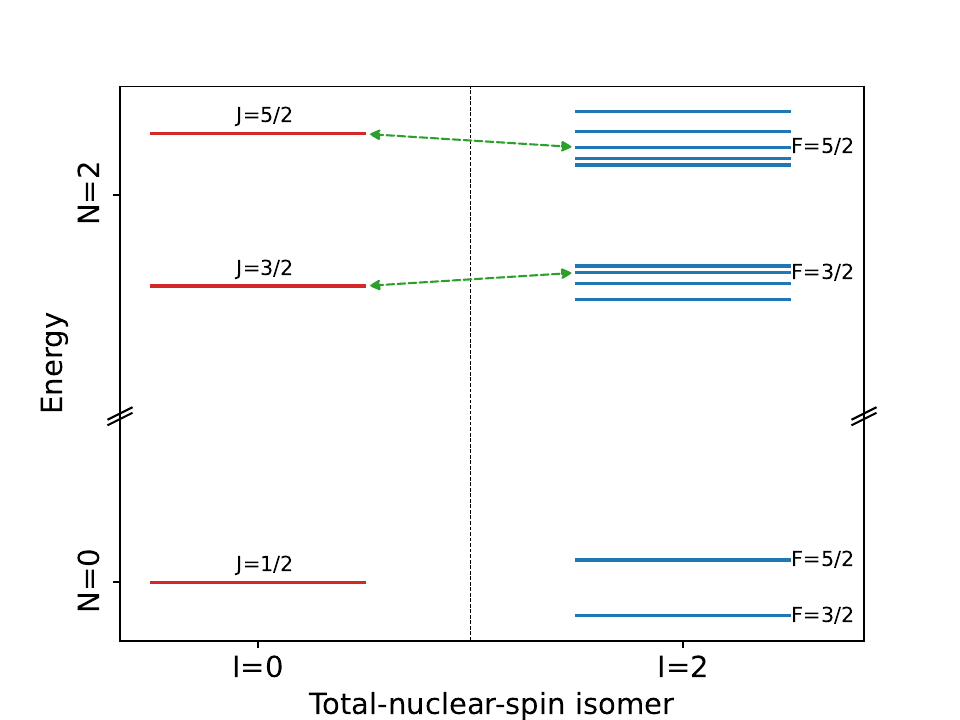}
    \caption{Energy level diagram of lowest rotational states ($N=0,2$) of the two ortho-NSIs ($I=0,2$) of $^{14}\textrm{N}_2^+$ molecular ion in the electronic ground state. The magnetic field is set to zero such that all Zeeman states coalesce. The green arrows indicate states that are coupled by the electric-quadrupole hyperfine interaction and can be tuned to degeneracy by applying an external magnetic field (see Fig. \ref{fig:crossing}). 
    }
    \label{fig:energies}
\end{figure}

Nuclear-spin isomers with the same nuclear-spin parity couple through the electric-quadrupole hyperfine interaction~\cite{bardeen1948calculation, brown2003rotational}. This coupling is typically much smaller than the magnetic hyperfine interaction of the nuclear spin with the electrons. Thus, the mixing of different NSI states generally is very small (e.g., in $\textrm{N}_2^+$ first excited rotational state ($N=2$), the mixing is $\mathcal{O}(10^{-5})$). Moreover, electromagnetic radiation doesn't couple strongly to the nuclear spin; hence, it doesn't mix different NSIs. For that, the different NSIs can be treated as distinct molecules with vastly different spectrums (Fig. \ref{fig:energies}).  

\begin{figure}
    \centering
    \includegraphics[width=\linewidth]{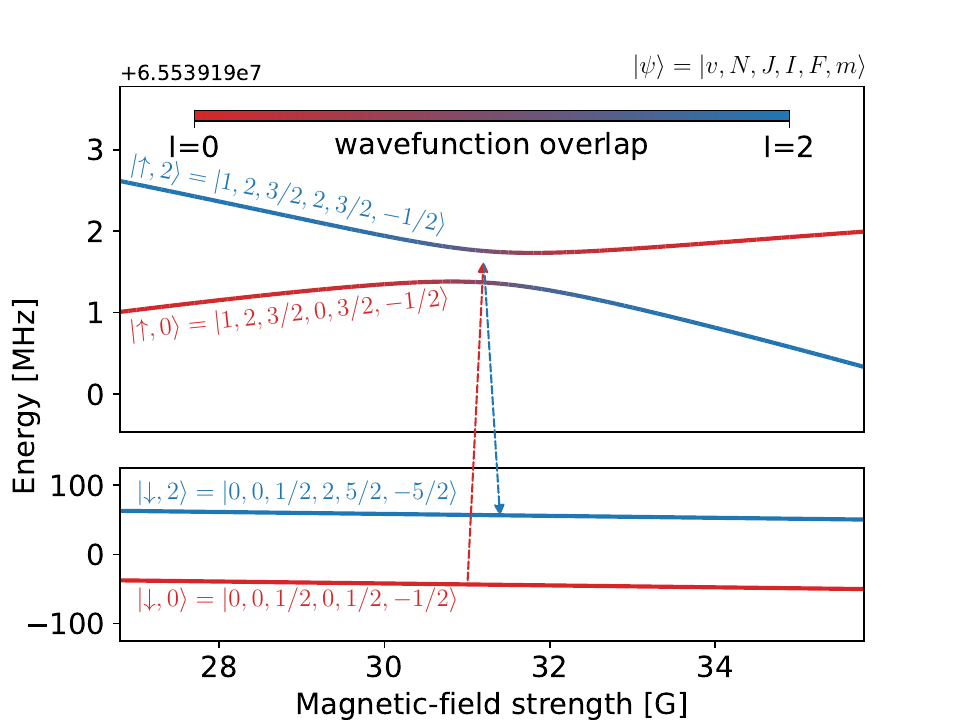}
    \caption{Adiabatic energy eigenstates of $\textrm{N}_2^+$ in the electronic ground state for finite external magnetic field. Top) Avoided crossing between two distinct NSI states (red - $I=0$, blue - $I=2$, Eq. \ref{eq:NSI_aux}). At $\sim31.2$ G, the eigenstates equally mix both NSI (legend). Bottom) Two pure NSI states that constitute the NSI qubit (Eq. \ref{eq:NSI_qubit}). Dashed arrows represent external fields that couple states of the same NSI quantum number, $I$. Wavefunction labels are: $v$ - vibration, $N$ - rotation, $J$ ($F$) - total angular momentum excluding (including) nuclear spin, $m$ - projection of the total angular momentum along the magnetic field axis. The electron spin quantum number, $S=1/2$, was omitted for brevity.  
    }
    \label{fig:crossing}
\end{figure}

As pointed out in Ref.~\cite{Najafian2020FromN2+}, we can enhance the mixing of two coupled NSI states by tuning the states' energies to degeneracy using an external magnetic field (Fig. \ref{fig:crossing}). At the avoided-crossing point, the energy eigenstates are the symmetric and antisymmetric superpositions of the NSI states. At this point, a gap appears in the spectrum, the size of which is twice the strength of the hyperfine electric-quadrupole coupling. This mixing opens a route for creating coherent superpositions of distinct NSI states and NSI-based qubits, as described in this paper.

Here, we describe in detail our scheme to create an NSI qubit between two non-mixed NSI states. We use a Raman-type coupling, exploiting NSIs mixing near the avoided crossing of two additional auxiliary states (Fig. \ref{fig:crossing}). We investigate the NSI-qubit dynamics by solving our system's four-level Hamiltonian model. We see an interference effect that governs the NSIs' effective-coupling strength as a function of laser detuning from the avoided crossing. We observe a crossover going from weak to strong coupling and the saturation of the NSIs' effective coupling by the strength of the electric-quadrupole hyperfine interaction. A sweet spot in the lasers' coupling strength occurs at the crossover location. We investigate the coupling-strength scaling as a function of the magnetic field, observing a Lorentzian-type decay once departing from the avoided-crossing point. Our calculations show the proposed scheme can achieve coherent manipulation and creation of qubits based on different NSIs.

\section{Experimental implementation in $\textrm{N}_2^+$}  

We follow Ref.~\cite{Najafian2020FromN2+} to find the eigenenergies and eigenstates of $\textrm{N}_2^+$ electronic ground state ($X\Sigma_g^+$) in a finite magnetic field. We use Hund's case ($b_{\beta_J}$) coupling scheme as a basis, 
\begin{equation*}
    \ket{\psi}=\ket{v,N,S,J,I,F,m},
\end{equation*}
where the quantum numbers are: $v$ for vibration, $N$ for rotation, $S$ for electron spin, $J$ for fine structure, $I$ for total nuclear spin, $F$ for hyperfine structure, and $m$ for total-angular-momentum projection along the magnetic field. We diagonalize the Hamiltonian (see Ref.~\cite{Najafian2020FromN2+}) in this basis using the set spanned by $I=0,2$ and $N=0,2,4$ for both the vibrational ground state ($v=0$) and the first excited state ($v=1$). 

The electric-quadrupole hyperfine interaction~\cite{mansour1991laser, Najafian2020FromN2+} couples the two ortho-NSI with equal $F$ quantum numbers (green arrows in Fig. \ref{fig:energies}). Applying a magnetic field can lead to an avoided crossing in the spectrum, as seen in Fig. \ref{fig:crossing}.

We choose the states,
\begin{align}\label{eq:NSI_qubit}
&\ket{\downarrow}_{Q}\equiv\down{I=0}\equiv\ket{0,0,1/2,1/2,0,1/2,-1/2},\\
\nonumber
&\ket{\uparrow}_{Q}\equiv\down{I=2}\equiv\ket{0,0,1/2,1/2,2,5/2,-5/2},
\end{align}
as the two NSI-qubit states. These two states have a definite NSI character as seen in Fig. \ref{fig:crossing}. Due to the two states' almost exact magnetic-field susceptibility~\cite{Najafian2020FromN2+}, these ``stretched'' states exhibit a very low relative magnetic-field susceptibility of $\sim0.6$ kHz/G, three orders of magnitude smaller than the bare susceptibility of the states. The qubit's reduced magnetic-field susceptibility makes it highly immune to decoherence induced by magnetic-field fluctuations~\cite{Najafian2020FromN2+}. 

We choose the states,
\begin{align}\label{eq:NSI_aux}
&\ket{\textrm{Aux}_0}\equiv\up{I=0}\equiv\ket{1,2,1/2,3/2,0,3/2,-1/2},\\
\nonumber
&\ket{\textrm{Aux}_2}\equiv\up{I=2}\equiv\ket{1,2,1/2,3/2,2,3/2,-1/2},
\end{align}
as the auxiliary states in our NSI-Raman scheme. These states exhibit an avoided crossing at a relatively low magnetic field of $\sim31.2$ G. The effective coupling (including geometrical factors) can be read directly from Fig. \ref{fig:crossing} as $\varepsilon\approx(2\pi)0.2$ MHz.

Our NSI-Raman scheme relies on coupling each arm of the NSI-qubit to the corresponding auxiliary state with the same NSI quantum number, $\down{I} \leftrightarrow \up{I}$. For our chosen states in $\textrm{N}_2^+$, the single-photon coupling corresponds to the dipole-forbidden S(0) transition at 4.5 $\mu$m. This mid-infrared electric-quadrupole transition was observed in Ref.~\cite{Germann2014}. A different path to couple the two states will be using a two-photon Raman scheme to drive rovibrational transitions in the electronic ground state in each NSI. The Raman beams are detuned from the dipole-allowed electronic excited state, $A^2\Pi_u$. Thus, in this scheme, the NSI qubit is driven by a total of four laser beams, all in the convenient near-infrared to optical regime.   


\section{Model Hamiltonian}
\begin{figure}
    \centering
    \includegraphics[width=\linewidth]{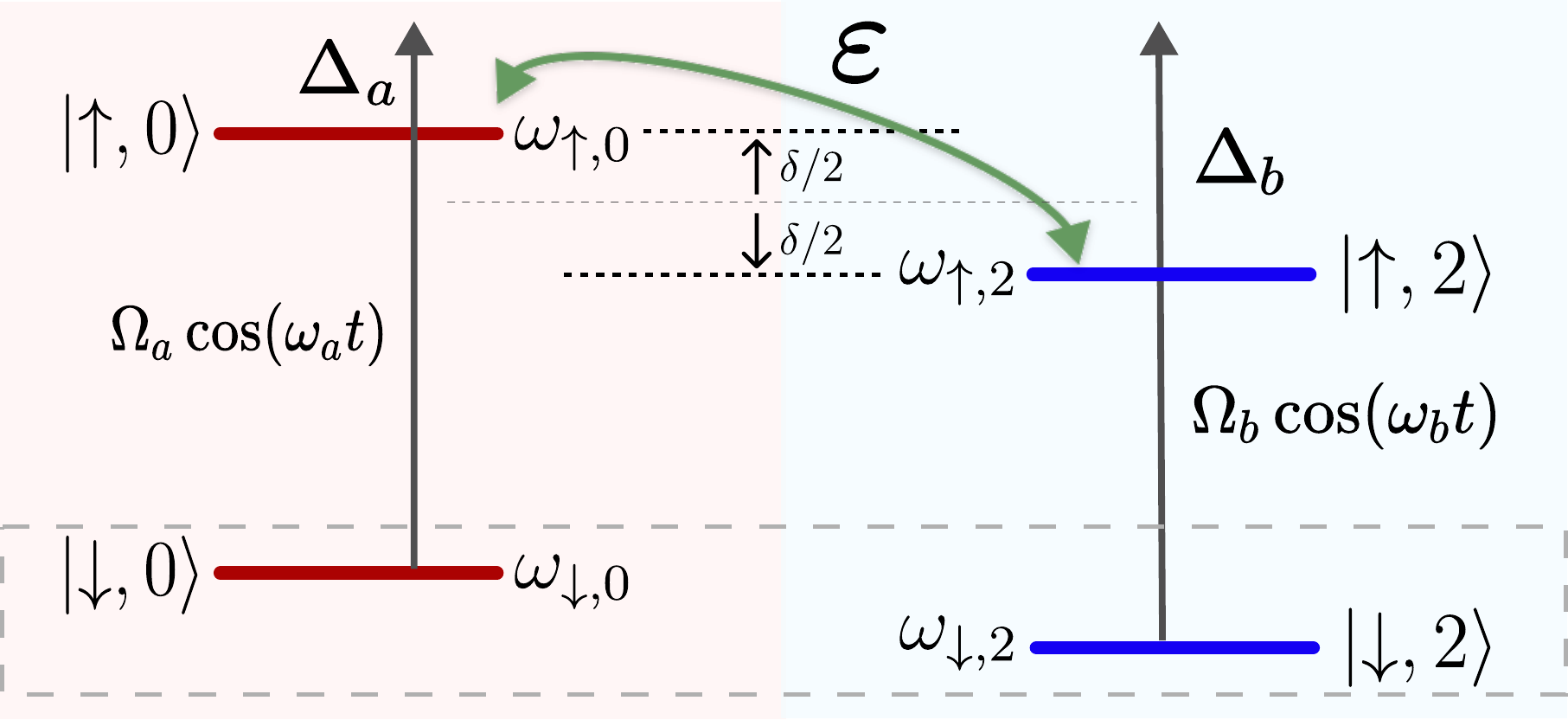}
    \caption{Four-level model for driving NSI qubits. Shaded red (blue) regions correspond to $I=0$ ($I=2$) NSI manifolds. The states $\down{0}$ and $\down{2}$ constitute the NSI qubit (dashed box). The states $\up{0}$ and $\up{2}$ were chosen due to their inherent coupling, $\varepsilon$, and magnetic-field-tunable energy splitting, $\delta$. The two levels in each manifold are coupled by external fields (thick black arrows) with coupling strength $\Omega_{a,b}$ and detuning from resonance, $\Delta_{a,b}$.}\label{fig:model}
\end{figure}

We model the NSI-qubit dynamics using a four-level Hamiltonian (see Fig. \ref{fig:model}),
\begin{multline}\label{eq:H_lab}
    H/\hbar=\\
    \left(
    \begin{array}{cc|cc}
        \omega_{\uparrow,0} & {\normalcolor \Omega_{a}\cos\left(\omega_{a}t\right)} & \varepsilon & 0\\
        \Omega_{a}\cos\left(\omega_{a}t\right) & \omega_{\downarrow,0} & 0 & 0\\
        \hline
        \varepsilon & 0 & \omega_{\uparrow,2} & \Omega_{b}\cos\left(\omega_{b}t\right)\\
        0 & 0 & \Omega_{b}\cos\left(\omega_{b}t\right) & \omega_{\downarrow,2}
    \end{array}
    \right),
\end{multline}
with a basis vector, $(\up{0},\down{0},\up{2},\down{2})^\intercal$. The lines in the Hamiltonian denote different NSI blocks. 
Here, we denote the NSI qubit states by $\down{0}$ and $\down{2}$, where both belong to the molecule's rovibrational ground state; however, they correspond to the $I=0$ and $I=2$ NSI, respectively (see Eq. \ref{eq:NSI_qubit}). The other two auxiliary levels are denoted by $\up{0}$ and $\up{2}$ (see Eq. \ref{eq:NSI_aux}). These levels are inherently coupled by the electric-quadrupole hyperfine interaction, $\varepsilon$. The corresponding state's energies (neglecting the effect of $\varepsilon$) are $\hbar\omega_{\downarrow,0}$, $\hbar\omega_{\downarrow,2}$, $\hbar\omega_{\uparrow,0}$, and $\hbar\omega_{\uparrow,2}$, respectively.
We apply two coupling fields simultaneously at $t=0$ that act separately on each nuclear-spin manifold, $\down{I} \leftrightarrow \up{I}$, with coupling strength $\Omega_a$ ($\Omega_b$) and frequency $\omega_a$ ($\omega_b$) for the $I=0$ ($I=2$) NSI. We omitted the initial phase of the coupling fields for brevity since it does not affect the results presented here~\cite{phase}. We assume that the state is initialized to $\down{0}$ at $t=0$.  

We transform this Hamiltonian into a rotated frame, $H_R=-i\hbar R \dot{R^\dagger}+RHR^\dagger$, using the unitary rotation matrix,
\begin{equation}\label{eq:RMatrix}
    R=
    e^{i(\omega_{\uparrow,2}+\omega_{\uparrow,0})t/2}
    \left(
    \begin{array}{cc|cc}
        1 & 0 & 0 & 0\\
        0 & e^{-i\omega_{a}t} & 0 & 0\\
        \hline
        0 & 0 & 1 & 0\\
        0 & 0 & 0 & e^{-i\omega_{b}t}
    \end{array}
    \right),
\end{equation}
and perform the rotating-wave approximation~\cite{marlan1997RWA} resulting in a time-independent Hamiltonian, 
\begin{equation}\label{eq:HRWA}
    H_{R}^\text{\tiny{RWA}}=\hbar\left(
    \begin{array}{cc|cc}
        \delta/2 & \Omega_{a}/2 & \varepsilon & 0\\
        \Omega_{a}/2 & \Delta_{a}+\delta/2 & 0 & 0\\
        \hline
        \varepsilon & 0 & -\delta/2 & \Omega_{b}/2\\
        0 & 0 & \Omega_{b}/2 & \Delta_{b}-\delta/2
    \end{array}
    \right).
\end{equation}
Here, $\Delta_{b}\equiv\omega_b-(\omega_{\uparrow,2}-\omega_{\downarrow,2})$ and  $\Delta_{a}\equiv\omega_{a}-(\omega_{\uparrow,0}-\omega_{\downarrow,0})$ are the detuning of each coupling field from its corresponding two-level resonance. The degeneracy of the auxiliary states is quantified by $\delta \equiv \omega_{\uparrow,0}-\omega_{\uparrow,2}$. The NSI-qubit resonant condition is given by $\Delta_b - \delta/2=\Delta_a+\delta/2\equiv\Delta$.

We solve this Hamiltonian's dynamics using numerical diagonalization. Further analytic approximations are given in the results section below.

\section{Results}
\subsection{Detuning, $\Delta$}

\begin{figure}
  \centering
  \includegraphics[width=\linewidth]{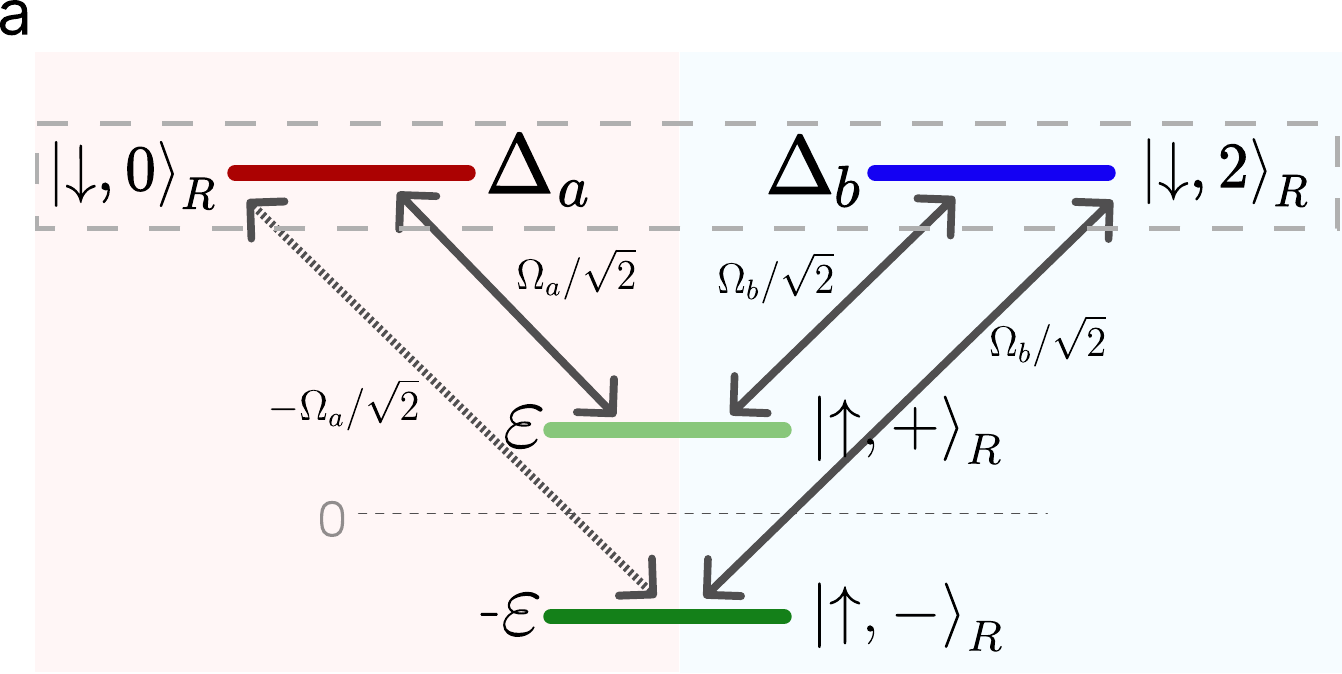}
  \vfill
  \includegraphics[width=\linewidth]{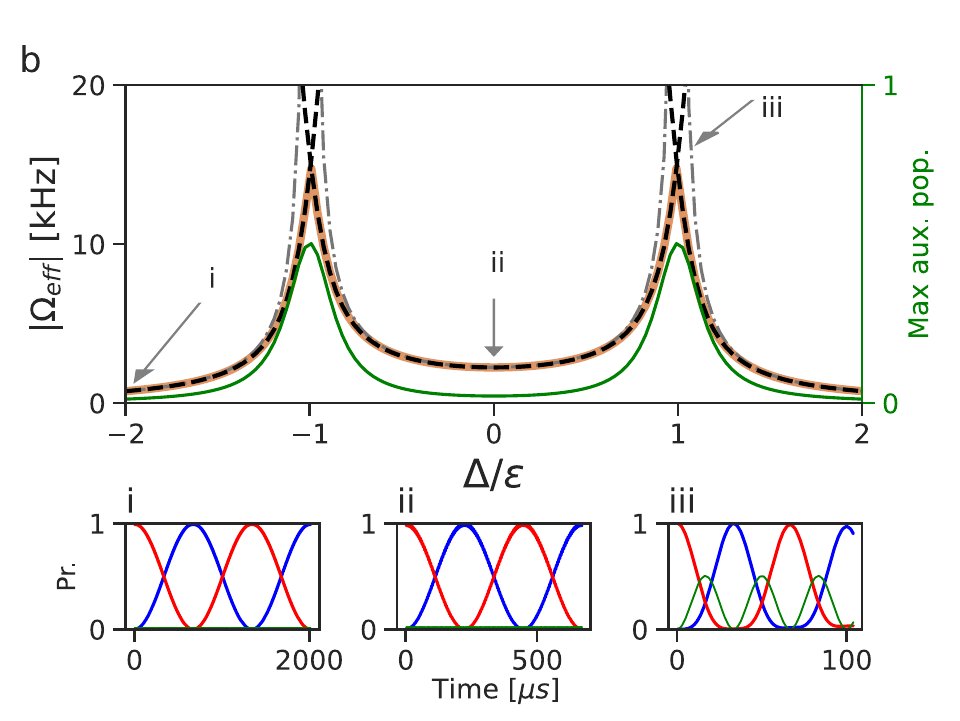}
  \caption{NSI-qubit dynamics for the degenerate ($\delta$ = 0) and resonant ($\Delta=\Delta_a=\Delta_b$) case. (a) The four-level model in the rotated frame (Eq. \ref{eq:H_rotated}) for this case. Shaded red (blue) regions correspond to $I=0$ ($I=2$) NSI manifolds. The grey dashed box indicates the NSI-qubit states. Green states mix both NSI manifolds. Arrows indicate the coupling between states by external fields (dotted arrow has a negative-sign coupling). (b) The effective-coupling magnitude as a function of the coupling-fields' detuning from the avoided-crossing energy. Here, $\Omega_a=\Omega_b=0.15\varepsilon=(2\pi)30$ kHz. The orange line is an exact numerical solution of Eq. \ref{eq:HRWA}, while the dash-dotted grey line is an analytical approximation using adiabatic elimination~\cite{Etienne20017Adiabatic,supp} (Eq. \ref{eq:Oeff_d0}). The dashed black line is an exact analytical solution~\cite{exact} of Eq. \ref{eq:H_rotated} (see also Eq. \ref{eq:H_rotated_both} and its related discussion). The green line is the maximal occupation in the auxiliary states, $\ket{\uparrow,\pm}$. (i)-(iii) Rabi oscillations for $\Delta/\varepsilon$ = -2,0,1, respectively. The red and blue lines are the probabilities of being in states $\ket{\downarrow,0}$ and $\ket{\downarrow,2}$, respectively, while the green line is the probability to be in the auxiliary states.}\label{fig:result_detuning}
\end{figure}

In Fig. \ref{fig:result_detuning}b, we show the effective NSI-qubit coupling, $\Omega_\text{eff}$, as a function of the coupling-fields' detuning where the auxiliary states are degenerate ($\delta=0$) and the NSI qubit is resonant ($\Delta=\Delta_a=\Delta_b$). We set the coupling-fields' strength to $\Omega_a=\Omega_b=0.15\varepsilon=(2\pi)30$ kHz (weak coupling).
We observe two resonant peaks at $\Delta=\pm\varepsilon$, a fast decay for $|\Delta|>\varepsilon$, and first-order constant coupling for $\Delta\sim0$.

To get further insight into the dynamic in these three regimes, we note that for the case of $\delta=0$, where the two inherently-coupled auxiliary states are degenerate, the eigenstates of the auxiliary manifold (without the coupling fields) are given by the symmetric and antisymmetric combination, $\ket{\uparrow,+}_R=(\up{2}_R+\up{0}_R)/\sqrt{2}$, and $\ket{\uparrow,-}_R=(\up{2}_R-\up{0}_R)/\sqrt{2}$ (see Fig. \ref{fig:result_detuning}a green levels). Here, the subscript $R$ denotes we are in the rotated frame, $\ket{\psi}_R=R\ket{\psi}$. In this basis, the Hamiltonian (including the coupling fields) is given by
\NiceMatrixOptions{cell-space-limits = 2pt}
\begin{equation}\label{eq:H_rotated}
    \tilde{H}_R^\text{\tiny{RWA}}=\hbar
    \begin{pNiceArray}{cc:cc}
        \varepsilon & 0 & \frac{\Omega_{b}}{2\sqrt{2}} & \frac{\Omega_{a}}{2\sqrt{2}}\\
        0 & -\varepsilon & \frac{\Omega_{b}}{2\sqrt{2}} & -\frac{\Omega_{a}}{2\sqrt{2}}\\
        \hdottedline
        \frac{\Omega_{b}}{2\sqrt{2}} & \frac{\Omega_{b}}{2\sqrt{2}} & \Delta_b & 0 \\
        \frac{\Omega_{a}}{2\sqrt{2}} & -\frac{\Omega_{a}}{2\sqrt{2}} & 0 & \Delta_{a}
    \end{pNiceArray},
\end{equation}
with a basis vector $(\ket{\uparrow,+}_R,\ket{\uparrow,-}_R,\down{2}_R,\down{0}_R)^\intercal$. The dotted lines in the Hamiltonian differentiate between the qubit subspace and the auxiliary subspace. 

We see that the NSI qubit states, $\down{0/2}_R$, are coupled by two paths going through the symmetric and antisymmetric auxiliary states (see Fig. \ref{fig:result_detuning}a black arrows). For the NSI qubit resonant case, $\Delta_a \sim \Delta_b$, when the auxiliary states are far-detuned, $|\Delta\pm\varepsilon|\gg\Omega_{a,b}$, we get an effective two-level system~\cite{Etienne20017Adiabatic} with coupling,
\begin{align}\label{eq:Oeff_d0}
    \Omega_\text{eff}&=
    \frac{\Omega_b\Omega_a}{4}\left(\frac{1}{\Delta-\varepsilon}-\frac{1}{\Delta+\varepsilon}\right).
\end{align}
This coherent two-path interference captures the resonant peaks and scaling observed in Fig. \ref{fig:result_detuning}b (dash-dotted grey line).

The effective detuning, 
\begin{align}\label{eq:Deff_d0}
    \Delta_\text{eff}&=
    \frac{\Omega_{a}^{2}-\Omega_{b}^{2}}{8}
    \left(\frac{1}{\Delta-\varepsilon}+\frac{1}{\Delta+\varepsilon} \right)
    +\left(\Delta_{a}-\Delta_{b}\right),
\end{align}
has contributions from the difference of the induced ac-Stark shifts on the two NSI qubit states. For $\Omega_a=\Omega_b$ and $\Delta_a=\Delta_b$, the effective detuning vanishes due to the symmetry of the Hamiltonian for $\delta=0$. In the supplemental material, we derive Eqs. \ref{eq:Oeff_d0} and \ref{eq:Deff_d0} using the adiabatic-elimination procedure~\cite{Etienne20017Adiabatic,supp}.

In the insets of Fig. \ref{fig:result_detuning}b, we show Rabi oscillations for different NSI-qubit detuning. For $\Delta/\varepsilon=-2,0$ (inset (i) and (ii) respectively), we observe oscillation between the NSI-qubit energy levels (red and blue lines) without almost any excitation of the auxiliary levels (green line). For $\Delta/\varepsilon=1$ (inset (iii)), we get the fastest oscillations; however, the auxiliary states participate in the dynamics. So, in this regime, our NSI-qubit operations are limited to inversion gates only (see Discussion). 

\subsection{Coupling-strength, $\Omega$}

\begin{figure}
  \includegraphics[width=\linewidth]{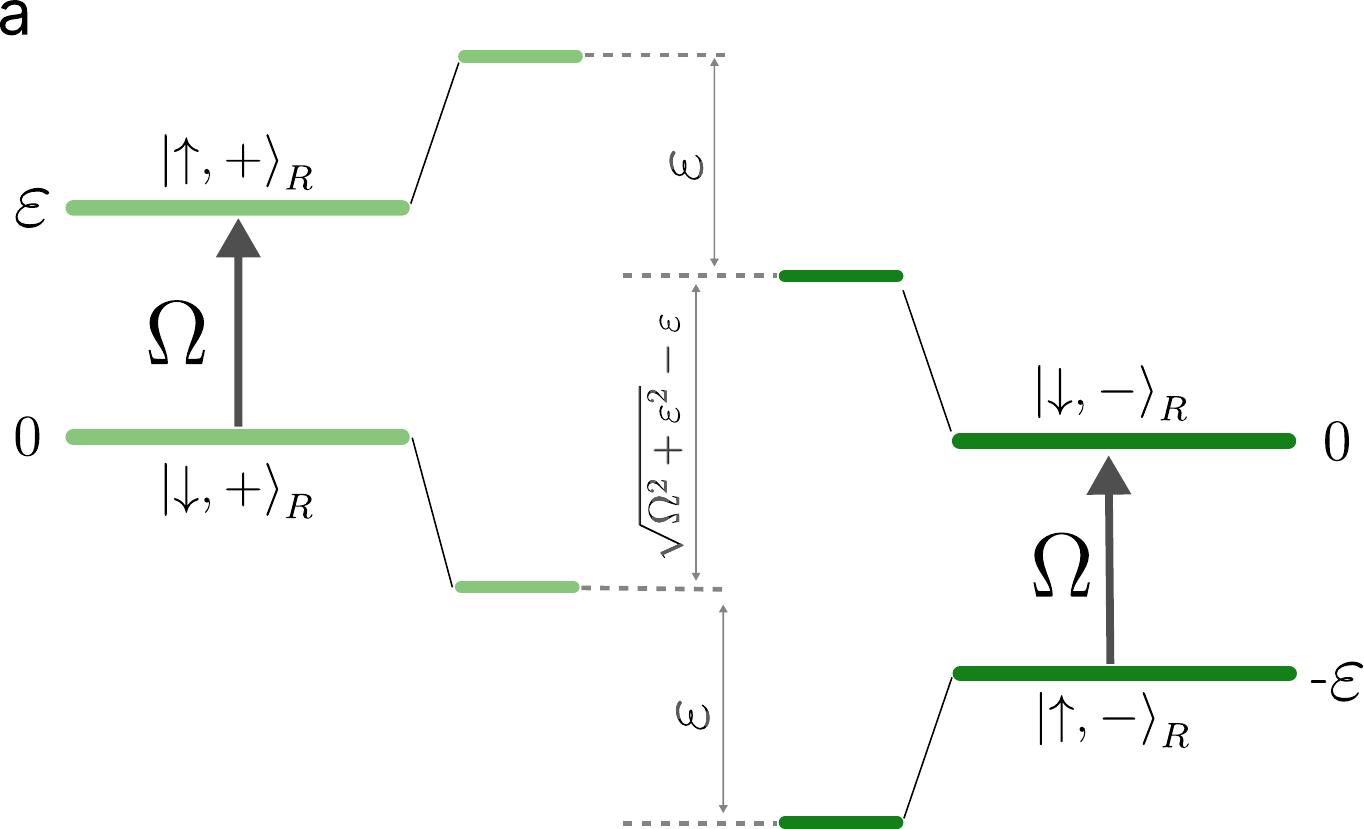}\label{fig:symmetric}
  \centering
  \includegraphics[width=\linewidth]{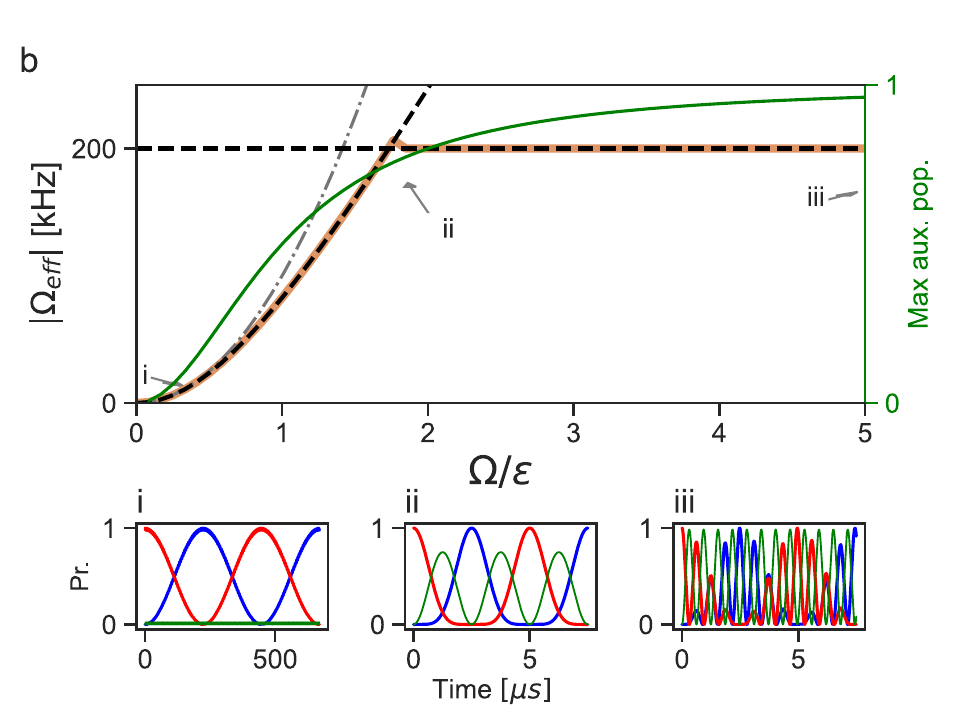}\label{fig:Detuning}
  \caption{NSI-qubit dynamics as a function of the coupling-fields' strength, $\Omega\equiv\Omega_a=\Omega_b$, for the resonant ($\Delta_a = \Delta_b = 0$) and degenerate ($\delta=0$) case. (a) Energy-level diagram for the Hamiltonian in Eq. \ref{eq:H_rotated_both}. Left (light green) and right (dark green) regions correspond to the symmetric and antisymmetric mixed NSI manifolds, respectively. Short lines are the ac-stark-shifted energy levels. Thick black arrows indicate the coupling between states by external fields, while thin grey arrows indicate energy gaps. (b) Effective-coupling magnitude as a function of the coupling-fields' strength. Here  $\varepsilon=(2\pi)200$ kHz. The orange and dashed black lines are the exact numerical and analytical~\cite{exact} solutions of the Hamiltonian in Eq. \ref{eq:H_rotated_both}, respectively. The dash-dotted grey line is an analytical approximation using adiabatic elimination~\cite{Etienne20017Adiabatic,supp} (Eq. \ref{eq:Oeff_d0}). The green line is the maximal occupation in the auxiliary states, $\ket{\uparrow,\pm}$. (i)-(iii) Rabi oscillations for $\Omega/ \varepsilon= 0.15,\sqrt{3},8$, respectively. The red and blue lines are the probabilities of being in states $\ket{\downarrow,0}$ and $\ket{\downarrow,2}$, respectively, while the green line is the probability of being in the auxiliary states.}\label{fig:result_omega}
\end{figure}

We now focus on the zero-detuning case, $\Delta=0$, where the contributions from the symmetric and antisymmetric auxiliary states interfere constructively. We increase the coupling strength from the weak ($\Omega<\varepsilon$) to the strong ($\Omega>\varepsilon$) regime. 
The results are shown in Fig. \ref{fig:result_omega}b. 
We see a crossover of the NSI-qubit effective coupling from the weak regime, $\Omega_\text{eff}\sim\Omega^2/\varepsilon$ (Eq. \ref{eq:Oeff_d0}), to the strong regime, $\Omega_\text{eff}\sim\varepsilon$, emphasizing that the strength of the electric-quadrupole hyperfine interaction bounds the maximal NSI-qubit coupling. 

To get insight into this crossover, we note that for equal-strength coupling fields, $\Omega_a=\Omega_b\equiv\Omega$, the Hamiltonian takes a straightforward block-diagonal form,
\begin{equation}\label{eq:H_rotated_both}
    \tilde{\tilde{H}}_R^\text{\tiny{RWA}}=\hbar
    \begin{pNiceArray}{cc||cc}
        \varepsilon & \Omega/2 & 0 & 0\\
        \Omega/2 & \Delta & 0 & 0\\
        \Hline\Hline
        0 & 0 & -\varepsilon & \Omega/2 \\
        0 & 0 & \Omega/2 & \Delta
    \end{pNiceArray},
\end{equation}
with a basis vector $(\ket{\uparrow,+}_R,\ket{\downarrow,+}_R,\ket{\uparrow,-}_R,\ket{\downarrow,-}_R)^\intercal$. The double lines in the Hamiltonian differentiate between the symmetric and antisymmetric subspaces of mixed NSIs.
The eigenvalues of this Hamiltonian (for $\Delta=0$) are given by $\pm\varepsilon/2\pm\sqrt{\varepsilon^2+\Omega^2}/2$ (Fig. \ref{fig:result_omega}a). We note that the dynamics are governed by the lowest transition frequency of the spectrum (see ac-Stark-shifted energy levels in Fig. \ref{fig:result_omega}a , black dashed lines in Fig. \ref{fig:result_omega}b, and~\cite{exact}). The maximal effective coupling of $\Omega_\text{eff}=\varepsilon$ is reached with coupling-fields' strength of $\Omega\ge\sqrt{3}\varepsilon$.

In the insets of Fig. \ref{fig:result_omega}b, we show Rabi oscillations for different couplings. For $\Omega/\varepsilon=0.15$ (weak regime, inset (i)), we get an effective two-level system with almost any excitation of the auxiliary states. For $\Omega/\varepsilon=\sqrt{3}$ (cross over, inset (ii)), we get a sweet spot where the effective coupling is maximized, and the population-oscillation frequency between the qubit and the auxiliary states is minimal. For stronger couplings (e.g., inset (iii), where $\Omega/\varepsilon=8$), the effective coupling is also maximized; however, we see faster oscillation between the qubit and auxiliary states. This will potentially reduce the fidelity of NSI inversion-gate operation.   

\subsection{Auxiliary-states' energy-gap, $\delta$}

\begin{figure}
  \centering
  \includegraphics[width=\linewidth]{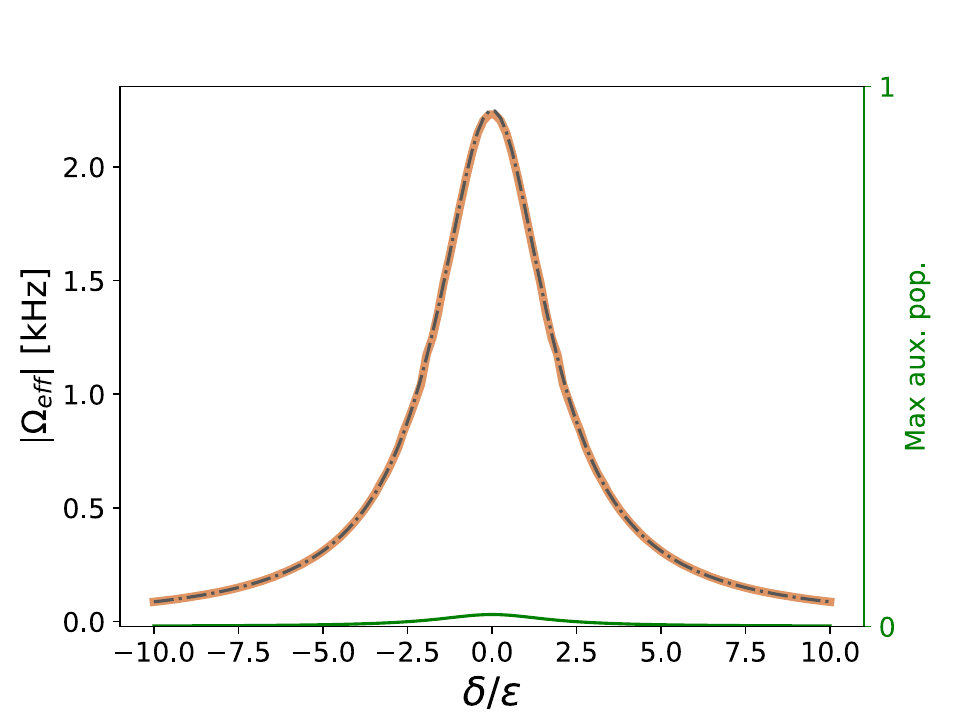}
  \caption{NSI-qubit effective coupling as a function of the auxiliary states' degeneracy, $\delta$. The coupling-fields' detunings are set to $\Delta_a=-\delta/2 - \xi/2$ and $\Delta_b=\delta/2 + \xi/2$. Here, $\xi$ is a correction due to the imbalanced ac-Stark shift of the two coupling fields such that $\Delta_\text{eff}=0$ (see Eq. \ref{eq:Deff}). The coupling-fields' amplitudes are set to $\Omega_a = \Omega_b = 0.15\varepsilon=(2\pi)30$ kHz. The orange line is an exact numerical solution of Eq. \ref{eq:HRWA}, while the dash-dotted grey line is an analytical approximation using adiabatic elimination (Eqs. \ref{eq:Oeff} and \ref{eq:Lorentzian}). The green line is the maximal occupation in the auxiliary states, $\ket{\uparrow,\pm}$.}\label{fig:result_degeneracy}
\end{figure}

In Fig. \ref{fig:result_degeneracy}, we expand the previous results, showing the effective NSI-qubit coupling (in the weak regime) as a function of the energy gap, $\delta$, between the auxiliary states (neglecting the hyperfine mixing).
Here, the eigenstates of the auxiliary states after mixing are given by $\ket{\uparrow,+_\theta}_R=\cos\theta\up{2}_R+\sin\theta\up{0}_R$ and $\ket{\uparrow,-_\theta}_R=\sin\theta\up{2}_R-\cos\theta\up{0}_R$, where the angle $\theta$ is linked to the energy gap of the auxiliary states by 
\begin{equation}
    \tan2\theta = 2\varepsilon/\delta.
\end{equation}
For the degenerate case, $\delta=0$, the mixing angle is $\theta=\pi/4$, while for the regime of negligible mixing, $\delta\gg\varepsilon$, we get $\theta\rightarrow0$.

The eigenenergies of the auxiliary states (neglecting coupling fields) are given by
\begin{equation}
    E_{\pm} = \pm \frac{\varepsilon}{\sin(2\theta)},
\end{equation}
and the Hamiltonian is given by (cf. Eq. \ref{eq:H_rotated}),
\begin{equation}\label{eq:H12}
    \begin{pNiceArray}{cc:cc}
       E_{+} & 0 & \frac{\Omega_{b}}{2}\cos\theta & \frac{\Omega_{a}}{2}\sin\theta\\
        0 & E_{-} & \frac{\Omega_{b}}{2}\sin\theta & -\frac{\Omega_{a}}{2}\cos\theta\\
        \hdottedline
        \frac{\Omega_{b}}{2}\cos\theta & \frac{\Omega_{b}}{2}\sin\theta & \Delta& 0\\
        \frac{\Omega_{a}}{2}\sin\theta & -\frac{\Omega_{a}}{2}\cos\theta & 0 & \Delta
    \end{pNiceArray},
\end{equation}
with a basis vector $(\ket{\uparrow,+_\theta}_R,\ket{\uparrow,-_\theta}_R,\down{2}_R,\down{0}_R)^\intercal$. 
The dotted lines in the Hamiltonian differentiate between the qubit subspace and the auxiliary subspace.
As in Fig. \ref{fig:result_detuning}a, a two-path coupling exists between the NSI qubit states through the two auxiliary states. 

When the auxiliary states are far detuned, we get an effective two-level system with coupling (in the adiabatic elimination approximation~\cite{Etienne20017Adiabatic,supp}), 
\begin{equation}\label{eq:Oeff}
    \Omega_\text{eff}=
    \frac{\Omega_{a}\Omega_{b}\sin\theta\cos\theta}{2}
    \left(\frac{1}{\Delta-E_{+}}
    -\frac{1}{\Delta-E_{-}}\right),
\end{equation}
and detuning, 
\begin{align}\label{eq:Deff}
    \nonumber
    \Delta_\text{eff}&=
    \frac{1}{4}\left(\frac{\Omega_{a}^{2}\sin^{2}\theta}{\Delta-E_{+}}-\frac{\Omega_{b}^{2}\cos^{2}\theta}{\Delta-E_{+}}+
    \frac{\Omega_{a}^{2}\cos^{2}\theta}{\Delta-E_{-}}-
    \frac{\Omega_{b}^{2}\sin^{2}\theta}{\Delta-E_{-}}\right)\\
    &+\left(\Delta_{a}+\delta-\Delta_{b}\right).
\end{align}
However, due to the asymmetry of this Hamiltonian, even for the resonant case, $\Delta = \Delta_a + \delta/2 = \Delta_b - \delta/2$, and equal coupling-fields' strength, $\Omega_a=\Omega_b$, the effective detuning is not zero. For that, we shift the relative coupling-fields' detuning, $\Delta_{b}-\Delta_{a}$, by $\Delta_\text{eff}$ to retrieve full NSI-qubit population inversion in the dynamics of Fig. \ref{fig:result_degeneracy}. The effective coupling is given by a Lorentzian (dashed grey line in Fig. \ref{fig:result_degeneracy}),
\begin{equation}\label{eq:Lorentzian}
    \Omega_\text{eff}=-\frac{\Omega_a\Omega_b}{2}\frac{\varepsilon}{(\delta/2)^2+\varepsilon^2}.
\end{equation}
Far from the avoided-crossing, $\delta\gg\varepsilon$, the effective coupling drops like $\Omega_\text{eff}\sim2\Omega_a\Omega_b\varepsilon/\delta^2$. This emphasizes the importance of working at a magnetic-field value where the auxiliary states are close to degeneracy ($\delta<\varepsilon$). 

\section{Discussion and summary}
In this paper, we have shown a general scheme to strongly couple two states originating from separate subspaces. We did so by coupling these states to two auxiliary states, one from each subspace, that inherently mix. The maximal coupling is limited to the strength of the auxiliary states' coupling, given that they can be tuned to degeneracy. The above level structure and degeneracy tuning occur naturally for the nitrogen molecular ion NSI degree of freedom. We have analyzed our scheme numerically and analytically for the relevant experimental parameters of this molecule. 

We have shown the dynamics of an effective two-level system for the NSI states, thus creating an effective NSI qubit. This two-level-system dynamic is limited to the weak-coupling regime, where the maximal effective-coupling strength is about two orders of magnitude below the mixing strength. In the strong coupling regime, the effective coupling strength can reach the mixing strength. However, in this regime, auxiliary states actively participate in the dynamics, breaking the two-level approximation. Nevertheless, we showed a fast inversion gate of the NSI states without populating the auxiliary states. The best parameters for this gate are given by, $\Omega=\sqrt{3}\varepsilon$, $\Delta=0$, and $\delta=0$.

We can use the NSI inversion gate with an additional resource to initialize any NSI superposition. 
For example~\cite{superposition}, by first creating a ``regular'' quantum superposition of a definite NSI character (e.g., $I=0$) and then using the inversion gate to invert one arm of the superposition (encoded in the $\ket{\downarrow,0}$) to the other isomeric state ($\ket{\downarrow,2}$ in this example). 
The creation of arbitrary superpositions of NSI states in the strong-coupling regime within the four-level model is left for future work.

We would like to thanks Alexander Poddubny, Nitzan Akerman, and Yuval Shagam for fruitful discussions and reading of this manuscript, and Shimon Levit for discussions on the hyperfine electric-quadrupole interaction in molecules.
We acknowledge the support of the Diane and Guilford Glazer Foundation Impact Grant for New Scientists, the Center for New Scientists at the Weizmann Institute of Science, the Edith and Nathan Goldenberg Career Development Chair, the Israel Science Foundation (1010/22), and the Minerva Stiftung with funding from the Federal German Ministry for Education and Research.

%

\newpage
~
\newpage
~
\onecolumngrid
\section{Supplemental material for coherent dynamics of a nuclear-spin-isomer superposition}

\subsection{Adiabatic elimination for degenerate ($\delta = 0$) auxiliary states}\label{appendix:degenerate}

We derive Eqs. \ref{eq:Oeff_d0} and \ref{eq:Deff_d0} of the main text using the adiabatic-elimination procedure \cite{Etienne20017Adiabatic}. To perform the adiabatic-elimination procedure, we choose a slightly different rotating frame, $R'=e^{-i\Delta t}R$, where $R$ is defined in Eq. \ref{eq:RMatrix} in the main text, and $\Delta \equiv (\Delta_a + \Delta_b)/2$ is a symmetric detuning. The effect of the additional global phase in the rotating matrix is to shift the diagonal terms in the rotated Hamiltonian (cf. Eq. \ref{eq:H_rotated} in the main text):   
\begin{equation}\label{eq:app:H_rotated}
    \tilde{H}_{R'}^\text{\tiny{RWA}}=\tilde{H}_{R}^\text{\tiny{RWA}}-\Delta I=\hbar
    \begin{pNiceArray}{cc:cc}
        \varepsilon - \Delta & 0 & \frac{\Omega_{b}}{2\sqrt{2}} & \frac{\Omega_{a}}{2\sqrt{2}}\\
        0 & -\varepsilon-\Delta & \frac{\Omega_{b}}{2\sqrt{2}} & -\frac{\Omega_{a}}{2\sqrt{2}}\\
        \hdottedline
        \frac{\Omega_{b}}{2\sqrt{2}} & \frac{\Omega_{b}}{2\sqrt{2}} & -\frac{\Delta_{ab}}{2} & 0 \\
        \frac{\Omega_{a}}{2\sqrt{2}} & -\frac{\Omega_{a}}{2\sqrt{2}} & 0 & \frac{\Delta_{ab}}{2}
    \end{pNiceArray},
\end{equation}
where $\Delta_{ab}\equiv\Delta_a-\Delta_b$. Schr\"{o}dinger's equation for the state $\ket{\psi}=a\ket{\uparrow,+}_R+b\ket{\uparrow,-}_R+c\down{2}_R+d\down{0}_R$ takes the form 
\begin{equation}
    \begin{cases}
        i\dot{a}\left(t\right)=\left(\varepsilon-\Delta\right)a+\frac{\Omega_{b}'}{2}c+\frac{\Omega_{a}'}{2}d\\
        i\dot{b}\left(t\right)=-\left(\varepsilon+\Delta\right)b+\frac{\Omega_{b}'}{2}c-\frac{\Omega_{a}'}{2}d\\
        i\dot{c}\left(t\right)=\frac{\Omega_{b}'}{2}a+\frac{\Omega_{b}'}{2}b-\frac{\Delta_{ab}}{2}c\\
        i\dot{d}\left(t\right)=\frac{\Omega_{a}'}{2}a-\frac{\Omega_{a}'}{2}b+\frac{\Delta_{ab}}{2}d\\
    \end{cases},
\end{equation}
where $\Omega'_{a,b}\equiv\Omega_{a,b}/\sqrt{2}$. 

Assuming $\left|\pm\varepsilon-\Delta\right|>>\Omega_{a,b},\left|\Delta_{ab}\right|$, we can perform the adiabatic-elimination approximation and set $\dot{a}\left(t\right)=\dot{b}\left(t\right)=0$, thus neglecting the dynamics of the auxiliary states, $\ket{\uparrow,\pm}$. Substituting the first two equations in the third and fourth equations gives us an effective two-level Hamiltonian,  
\begin{equation}
H_\text{eff} = \hbar
    \begin{pNiceArray}{cc}
         \Delta_\text{eff}^b & \Omega_\text{eff}/2\\
        \Omega_\text{eff}/2  & \Delta_\text{eff}^a\\    
    \end{pNiceArray},
\end{equation}
for the NSI qubit state, $\ket{\psi_\text{eff}}=c\down{2}_R+d\down{0}_R$, where 
\begin{equation}\label{eq:app:Oeff_general}
    \Omega_\text{eff}=
    \frac{\Omega_{a}\Omega_{b}}{4}
    \left(\frac{1}{\Delta-\varepsilon}
    -\frac{1}{\Delta+\varepsilon}\right)
\end{equation}
is Eq. \ref{eq:Deff_d0} of the main text, and 
\begin{align}\label{eq:app:Deff_d0}
    \Delta_\text{eff}\equiv\Delta_\text{eff}^a-\Delta_\text{eff}^b=
    \frac{\Omega_{a}^{2}-\Omega_{b}^{2}}{8}
    \left(\frac{1}{\Delta-\varepsilon}+\frac{1}{\Delta+\varepsilon} \right)
    +\left(\Delta_{a}-\Delta_{b}\right)
\end{align}
is Eq. \ref{eq:Oeff_d0} of the main text.

The above results are an example of a generalized Raman interaction with several auxiliary states. The effective Raman coupling for this case is given by a sum over all Raman interactions,
\begin{equation}
    \Omega_\text{eff}=\sum_i\frac{\Omega_{gi}\Omega_{if}}{2\Delta_i},
\end{equation}
and the effective Raman detuning is given by the difference in the ac-Stark shift between the two qubit states,
\begin{equation}
    \Delta_\text{eff}=\sum_i\frac{\Omega_{gi}^2}{4\Delta_i}-\sum_i\frac{\Omega_{if}^2}{4\Delta_i}.
\end{equation}

\subsection{Adiabatic elimination for the general case}\label{eq:app:general}
In the general case where the auxiliary states are not degenerate, $\delta\neq0$, the Hamiltonian in the rotated frame (using R' as in the previous section) is given by (cf. Eq. \ref{eq:H12} of the main text)
\begin{equation}
    \tilde{H}_{R'}^\text{\tiny{RWA}}=\hbar
    \begin{pNiceArray}{cc:cc}
       E_{+} - \Delta & 0 & \frac{\Omega_{b}}{2}\cos\theta & \frac{\Omega_{a}}{2}\sin\theta\\
        0 & E_{-} - \Delta& \frac{\Omega_{b}}{2}\sin\theta & -\frac{\Omega_{a}}{2}\cos\theta\\
        \hdottedline
        \frac{\Omega_{b}}{2}\cos\theta & \frac{\Omega_{b}}{2}\sin\theta & -\frac{\Delta_{ab}}{2}-\frac{\delta}{2} & 0\\
        \frac{\Omega_{a}}{2}\sin\theta & -\frac{\Omega_{a}}{2}\cos\theta & 0 & \frac{\Delta_{ab}}{2}+\frac{\delta}{2}
    \end{pNiceArray}.
\end{equation}
Here, $\Delta \equiv (\Delta_a+\Delta_b)/2$ as before. Similarly to the previous section, assuming $\left|E_{\pm}-\Delta\right|>>\Omega_{a,b},\left|\Delta_{ab}\pm\delta\right|$, we can eliminate the auxiliary stats. Following the procedure described above, we retrieve Eqs. \ref{eq:Oeff} and \ref{eq:Deff} of the main text. 

\subsection{Exact solutions for the resonant ($\Delta\equiv\Delta_a = \Delta_b$), degenerate ($\delta = 0$), and equal strength ($\Omega\equiv\Omega_a=\Omega_b$) case}
\begin{figure}
     \includegraphics[width=1.0\linewidth]{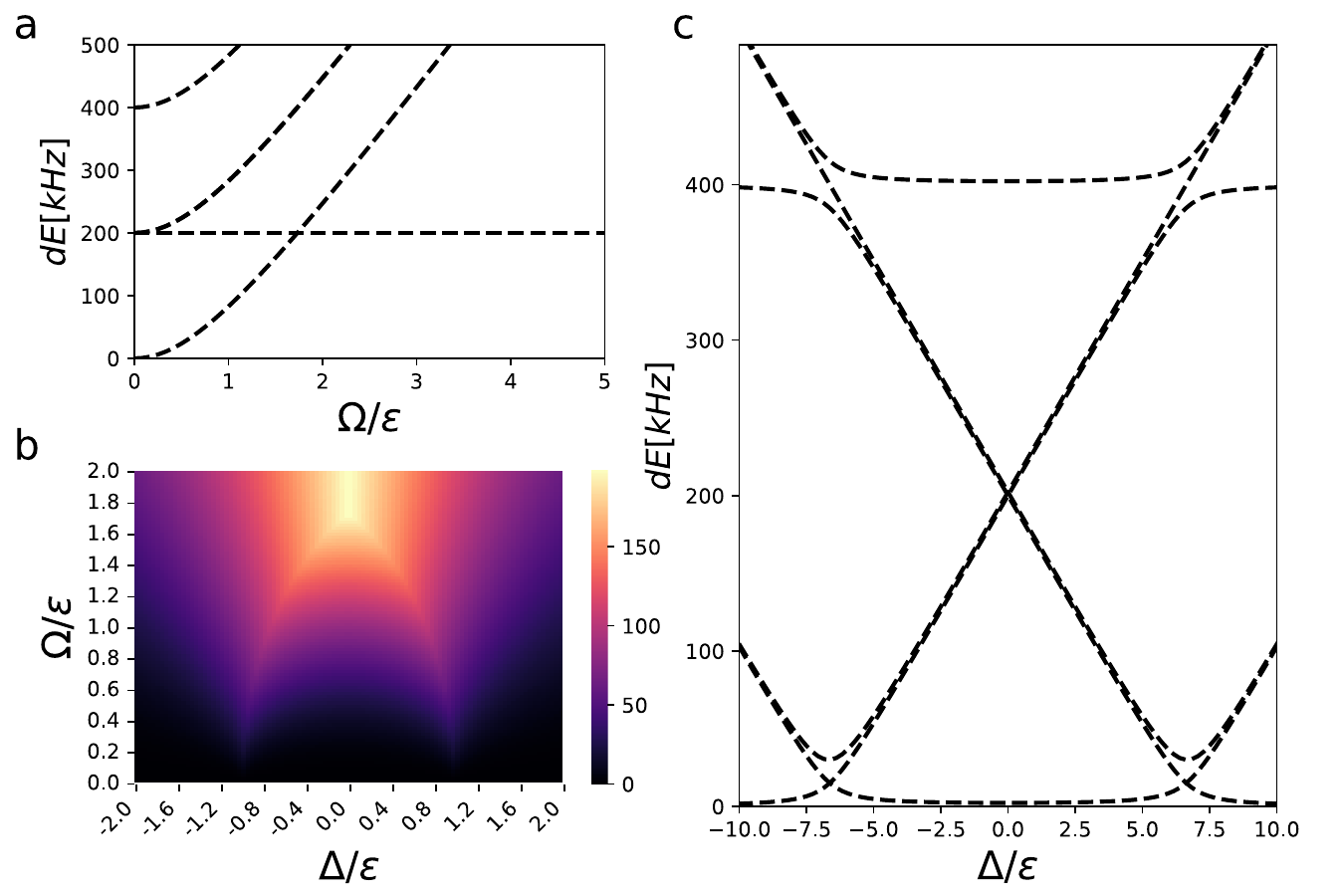}
     \caption{The spectrum's transition frequencies for the equal strength ($\Omega\equiv\Omega_a=\Omega_b$), resonant ($\Delta\equiv\Delta_a = \Delta_b$), and degenerate ($\delta=0$) case. The NSI-qubit coupling strength follows the lowest transition frequency in the spectrum. (a) Spectrum's transition frequencies as a function of the coupling-fields' strength, $\Omega$, for zero detuning ($\Delta = 0$). (b) The lowest transition frequency in the spectrum (the effective coupling strength of the NSI qubit) as a function of the coupling-fields' strength, $\Omega$, and the detuning, $\Delta$. (c) Spectrum's transition frequencies as a function of the detuning, $\Delta$, for $\Omega=0.15\varepsilon$.} \label{fig:app:spectrum}
\end{figure}

In Fig. \ref{fig:result_omega} and its related discussion in the main text, we give an exact solution (dashed black lines in Fig. \ref{fig:result_omega}b) for the effective dynamics for the case of zero detuning, $\Delta=0$. However, we can also derive an exact solution for the case of general detuning, $\Delta$, as can be seen in Fig. \ref{fig:result_detuning}b dashed black line in the main text. Solving analytically the time-independent Schr\"{o}dinger's equation for the Hamiltonian given in Eq. \ref{eq:H_rotated_both} of the main text 
, we get the following energy eigenvalues: 
\begin{equation}
    \begin{cases}
        \frac{1}{2}(\Delta-\varepsilon)\pm\frac{1}{2}\sqrt{\Omega^2+\left(\varepsilon+\Delta\right)^2}\\
        \frac{1}{2}(\Delta+\varepsilon)\pm\frac{1}{2}\sqrt{\Omega^2+\left(\varepsilon-\Delta\right)^2}
    \end{cases}.
\end{equation}
The spectrum's transition frequencies are given by 
\begin{equation}\label{eq:app:spectrum}
    \begin{cases}
        \sqrt{\Omega^2+\left(\varepsilon-\Delta\right)^2}\\
        \sqrt{\Omega^2+\left(\varepsilon+\Delta\right)^2}\\
        \varepsilon+\frac{1}{2}\sqrt{\Omega^{2}+\left(\varepsilon-\Delta\right)^{2}}-\frac{1}{2}\sqrt{\Omega^{2}+\left(\varepsilon+\Delta\right)^{2}}\\
        \varepsilon-\frac{1}{2}\sqrt{\Omega^{2}+\left(\varepsilon-\Delta\right)^{2}}-\frac{1}{2}\sqrt{\Omega^{2}+\left(\varepsilon+\Delta\right)^{2}}\\
        \varepsilon-\frac{1}{2}\sqrt{\Omega^{2}+\left(\varepsilon-\Delta\right)^{2}}+\frac{1}{2}\sqrt{\Omega^{2}+\left(\varepsilon+\Delta\right)^{2}}\\
        \varepsilon+\frac{1}{2}\sqrt{\Omega^{2}+\left(\varepsilon-\Delta\right)^{2}}+\frac{1}{2}\sqrt{\Omega^{2}+\left(\varepsilon+\Delta\right)^{2}}.
    \end{cases}
\end{equation}
In Fig. \ref{fig:app:spectrum} of this Supplemental Material, we plot the spectrum's transition frequencies as a function of the coupling strength for zero detuning (Fig. \ref{fig:app:spectrum}a) and as a function of the detuning for $\Omega=0.15\varepsilon$ (Fig. \ref{fig:app:spectrum}c). Comparing to Figs. \ref{fig:result_omega}b and \ref{fig:result_detuning}b of the main text, we see that the lowest transition frequency of the spectrum governs the NSI-qubit dynamics. In Fig. \ref{fig:app:spectrum}b, we give the analytic result for general detuning and coupling-field's strength. 

In the weak-coupling regime ($\Omega<\varepsilon$), the lowest transition frequency in the spectrum is given by (see Fig. \ref{fig:app:spectrum}c)

\begin{equation}
    \left|\varepsilon-\frac{1}{2}\sqrt{\Omega^{2}+\left(\varepsilon-\Delta\right)^{2}}-\frac{1}{2}\sqrt{\Omega^{2}+\left(\varepsilon+\Delta\right)^{2}}\right|,
\end{equation}
which is the equation for the dashed black line in Fig. \ref{fig:result_detuning}b. For zero detuning ($\Delta=0$), the lowest transition frequency in the spectrum exhibits a crossover at a coupling strength of $\Omega=\sqrt{3}\varepsilon$ (see Fig. \ref{fig:app:spectrum}a), 
\begin{equation}
    \left|\varepsilon-\sqrt{\Omega^{2}+\varepsilon^{2}}\right|
    \rightarrow\varepsilon 
    ,
\end{equation}
which is the equation for the dashed black line in Fig. \ref{fig:result_omega}b.

\newpage
\subsection{Phase-shifted coupling fields}
In Eq. \ref{eq:H_lab} of the main text, we assumed a particular phase (zero) for the coupling fields. Here, we show that our results remain the same for any choice of the initial phases, $\phi_{a,b}$, of the two coupling fields. We rewrite the model Hamiltonian (cf. Eq. \ref{eq:H_lab}) with additional phase terms to the coupling fields,
\begin{equation}\label{eq:app:H_lab_phase}
\centering
    H/\hbar=
    \left(
    \begin{array}{cc|cc}
        \omega_{\uparrow,0} & {\normalcolor \Omega_{a}\cos\left(\omega_{a}t-\phi_a\right)} & \varepsilon & 0\\
        \Omega_{a}\cos\left(\omega_{a}t-\phi_a\right) & \omega_{\downarrow,0} & 0 & 0\\
        \hline
        \varepsilon & 0 & \omega_{\uparrow,2} & \Omega_{b}\cos\left(\omega_{b}t-\phi_b\right)\\
        0 & 0 & \Omega_{b}\cos\left(\omega_{b}t-\phi_b\right) & \omega_{\downarrow,2}
    \end{array}
    \right),
\end{equation}
where $\Omega_{a,b}$ are real-valued as before.
Moving to the rotating frame and performing the rotating-wave approximation, the Hamiltonian is now changed to (cf. Eq. \ref{eq:H_rotated})
\begin{equation}
    H_{R}^\text{\tiny{RWA}}=\hbar\left(
    \begin{array}{cc|cc}
        \delta/2 & \Omega_{a}e^{+ i\phi_a}/2 & \varepsilon & 0\\
        \Omega_{a}e^{-i\phi_a}/2 & \Delta_{a}+\delta/2 & 0 & 0\\
        \hline
        \varepsilon & 0 & -\delta/2 & \Omega_{b}e^{+i\phi_b}/2\\
        0 & 0 & \Omega_{b}e^{-i\phi_b}/2 & \Delta_{b}-\delta/2
    \end{array}
    \right).
\end{equation}
Following the diagonalization procedure for the auxiliary states, we get the following Hamiltonian (cf. Eq. \ref{eq:H12}),
\begin{equation}
    H_{R}^\text{\tiny{RWA}}=\hbar
    \begin{pNiceArray}{cc:cc}
       E_{+} & 0 & \frac{\Omega_{b}}{2}e^{+i\phi_b}\cos\theta & \frac{\Omega_{a}}{2}e^{+i\phi_a}\sin\theta\\
        0 & E_{-} & \frac{\Omega_{b}}{2}e^{+i\phi_b}\sin\theta & -\frac{\Omega_{a}}{2}e^{+i\phi_a}\cos\theta\\
        \hdottedline
        \frac{\Omega_{b}}{2}e^{-i\phi_b}\cos\theta & \frac{\Omega_{b}}{2}e^{-i\phi_b}\sin\theta & \Delta& 0\\
        \frac{\Omega_{a}}{2}e^{-i\phi_a}\sin\theta & -\frac{\Omega_{a}}{2}e^{-i\phi_a}\cos\theta & 0 & \Delta
    \end{pNiceArray}.
\end{equation}
Using the adiabatic-elimination method as explained above, results in the effective coupling strength 
\begin{equation}\label{eq:app:Oeff31}
    \Omega^{\phi}_\text{eff}=
    \frac{\Omega_{a}\Omega_{b}e^{i(\phi_a-\phi_b)}\sin\theta\cos\theta}{4}
    \left(\frac{1}{\Delta-E_{+}}
    -\frac{1}{\Delta-E_{-}}\right).
\end{equation} 
Comparing both analytical approximation and exact numeric results, we see that $\left|\Omega^{\phi}_\text{eff}\right|$ = $\left|\Omega_\text{eff}\right|$ and $\Delta^{\phi}_\text{eff}$ =  $\Delta_\text{eff}$ such that the system's dynamics won't be affected by $\phi_{a,b}$ for the scheme presented here. For more elaborated schemes involving several pulses, such as Ramsey-type schemes, the relative phase between the pulses will affect the dynamics. 

\subsection{General NSI superpositions in the strong-coupling regime}

Due to the population of the auxiliary states in the strong-coupling regime, we are limited to performing only inversion gates on the NSI qubit. We define a general rotation matrix with a rotation vector along the equatorial of a two-level Bloch sphere:
\begin{equation}\label{eq:rotation}
    \mathcal{R}(\theta)=
    \begin{pNiceArray}{cc}
        \cos(\theta/2) & i\sin(\theta/2) \\
        i\sin(\theta/2) & \cos(\theta/2)
    \end{pNiceArray}.
\end{equation}
Using the above definition, the NSI inversion gate is given by $\mathcal{R}(\pi)$.

Adding an additional resource is one route to initialize an arbitrary superposition using only the inversion gate. We assume we can couple the $\ket{\downarrow,0}$ state with an additional state, the state $\ket{a,0}$. These states have the same NSI character; thus, they constitute a ``regular'' qubit. We assume we can perform operations between $\ket{\downarrow,0}$ and $\ket{a,0}$ without interrupting the other states in our four-level model Hamiltonian.

\begin{figure}
     \includegraphics[width=1.0\linewidth]{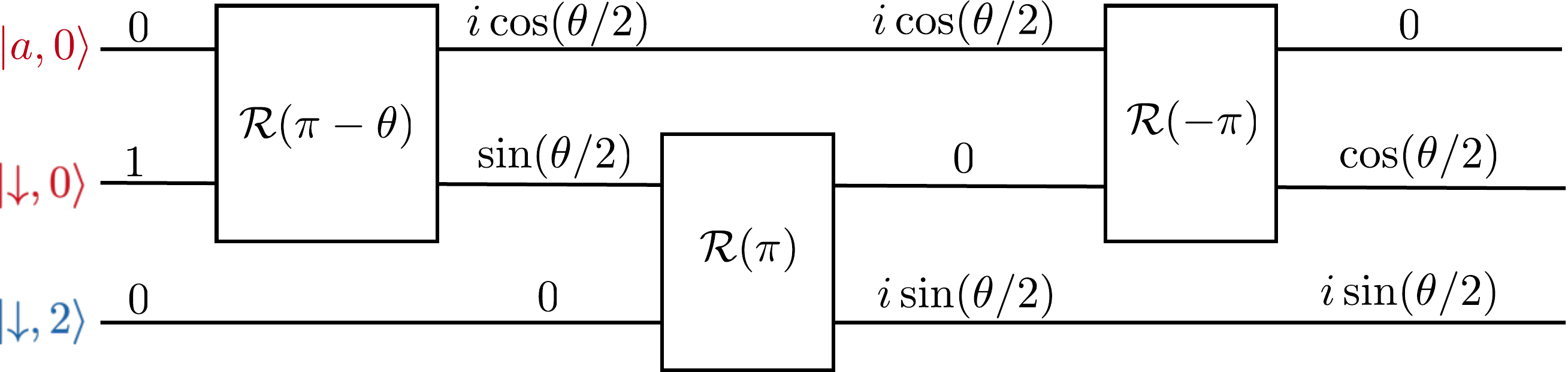}
     \caption{A protocol to initialize a general NSI superposition in the strong-coupling regime. Each line in the circuit represents a state. Each box represents a general rotation gate (Eq. \ref{eq:rotation}). The text above the lines is the amplitude of each state during the protocol (Eqs. \ref{eq:protocol1}-\ref{eq:protocol3}).}\label{fig:scheme}
 \end{figure}

In the following, we will use the basis $(\ket{a,0},\ket{\downarrow,0},\ket{\downarrow,2})^\intercal$. We start our protocol by ``shelving'' the part of the superposition we wish to keep in the $I=0$ manifold to the $\ket{a,0}$ state using an $\mathcal{R}(\pi-\theta)$ rotation,
\begin{equation}\label{eq:protocol1}
    \begin{pNiceArray}{ccc}
        \sin(\theta/2) & i\cos(\theta/2) & 0 \\
        i\cos(\theta/2) & \sin(\theta/2) & 0 \\
        0 & 0 & 1
    \end{pNiceArray}
    \begin{pNiceArray}{c}
        0\\
        1\\
        0
    \end{pNiceArray}
    =
    \begin{pNiceArray}{c}
        i\cos(\theta/2)\\
        \sin(\theta/2)\\
        0
    \end{pNiceArray}.
\end{equation}
Next, we use the NSI inversion gate, $\mathcal{R}(\pi)$, to transfer the remaining part of the superposition in the $\ket{\downarrow,0}$ to the $\ket{\downarrow,2}$ state,
\begin{equation}\label{eq:protocol2}
    \begin{pNiceArray}{ccc}
        1 & 0 & 0 \\
        0 & 0 & i \\
        0 & i & 0
    \end{pNiceArray}
    \begin{pNiceArray}{c}
        i\cos(\theta/2)\\
        \sin(\theta/2)\\
        0
    \end{pNiceArray}
    =
    \begin{pNiceArray}{c}
        i\cos(\theta/2)\\
        0\\
        i\sin(\theta/2)
    \end{pNiceArray}.
\end{equation}
Last, we ``deshelve'' the population in the $\ket{a,0}$ state back to the $\ket{\downarrow,0}$ state using an $\mathcal{R}(-\pi)$ rotation,
\begin{equation}\label{eq:protocol3}
    \begin{pNiceArray}{ccc}
        0 & -i & 0 \\
        -i & 0 & 0 \\
        0 & 0 & 1
    \end{pNiceArray}
    \begin{pNiceArray}{c}
        i\cos(\theta/2)\\
        0\\
        i\sin(\theta/2)
    \end{pNiceArray}
    =
    \begin{pNiceArray}{c}
        0\\
        \cos(\theta/2)\\
        i\sin(\theta/2)
    \end{pNiceArray}.
\end{equation}
In total, we initialized a general NSI superposition, $\cos(\theta/2)\ket{\downarrow,0}+i\sin(\theta/2)\ket{\downarrow,2}$. The above protocol is visualized in Fig. \ref{fig:scheme}.

\end{document}